\documentclass[twocolumn,aps,nofootinbib]{revtex4-1}
\pdfoutput=1
\usepackage{booktabs}
\usepackage{xcolor}
\usepackage[utf8]{inputenc}
\usepackage[english]{babel}

\usepackage{hyperref}
\usepackage[style=base]{caption}
\usepackage{multirow}
\usepackage{setspace}
\usepackage{amsmath, amssymb, amsthm}
\usepackage{cleveref}
\usepackage{physics}
\usepackage{url}
\usepackage{array}
\usepackage{tcolorbox}

\usepackage{graphicx,caption}

\begin{document}
\date{\today}
\author{Jacob L.L.C.C. Janssen$^{1}$} 
\email{Corresponding author.  Email: jacob.janssen@tno.nl.}
\author{Yaneer Bar-Yam$^2$}
\affiliation{$^1$ Amsterdam, TNO, the Netherlands}
\affiliation{$^2$New England Complex Systems Institute, Cambridge, MA}
\title{Lowest-cost virus suppression}

\begin{abstract}
Analysis of policies for managing epidemics require simultaneously an economic and epidemiological perspective. We adopt a cost-of-policy framework to model both the virus spread and the cost of handling the pandemic. Because it is harder and more costly to fight the pandemic when the circulation is higher, we find that the optimal policy is to go to zero or near-zero case numbers. Without imported cases, if  a  region is willing  to  implement  measures  to prevent spread  at  one  level  in  number  of  cases, it  must also be willing to prevent the spread with at a lower level, since  it  will  be cheaper  to  do  so  and  has  only  positive other effects. With imported cases, if a region is not coordinating with other regions, we show the cheapest policy is continually low but nonzero cases due to decreasing cost of halting imported cases. When it is coordinating, zero is cost-optimal. Our analysis indicates that within Europe cooperation targeting a reduction of both within country transmission, and between country importation risk, should help achieve lower transmission and reduced costs. 

\end{abstract}

\maketitle

\flushbottom 



\thispagestyle{empty} 





COVID-19, caused by the novel coronavirus SARS-CoV-2 has become a severe pandemic~\cite{WHO2020}. Understanding how the cost of suppressing the virus interacts with the dynamics of disease transmission, both within and between regions provides insight into the best course of action~\cite{acemoglu2020,bognana2020,bognana2020,kissler2020,siegenfeld2020,norman2020}. In this article, we show that policies that allow for a continually low virus circulation are optimal in situations both with and without inter-regional travel. Not considering the cost of limiting travel at first, we find the cost of suppressing the virus can be described as an increasing function of case numbers. Conversely, the cost of limiting imported cases is decreasing in the number of cases. When we abstract from dealing with inter-regional travel, the increasing nature of the cost function implies the cost of suppressing the virus is lowest at zero cases. Since there are only positive side effects to having less virus, the optimal policy is to go to zero. 

Due to globalization and inter-regional travel the world has become more vulnerable to pandemics and suppressing the virus has become more difficult~\cite{baryam2016,petersen2018} and more costly~\cite{antras2020}, sometimes even to the point of being used as an argument for giving up preventing the virus from becoming widespread entirely~\cite{Tegnell2020,GBD2020,rijksoverheid2020}. The cost of limiting imported cases when analysing multiple regions is based on the simultaneous cost of suppressing the virus within a region and the cost of limiting imported cases from outside the region. As the cost of limiting imported cases is decreasing in the number of cases, and the cost of suppressing the virus is increasing in the number of cases, a natural question is at what number of cases the cost of suppressing the virus is lowest. We show that in spite of the cost of limiting inter-regional travel, the impact of imported cases on costly society-wide restrictions implies the best course of action is to go to very low number of cases that can be controlled through testing, tracing and isolating (TTI) of individual cases.

\section{Costs of policy: Increasing in case numbers}

Naturally, when there is more virus, one must do more to stop the spread. This means the cost of suppressing the virus is increasing in the number of cases. Measures to halt the spread of the virus each have their own financial and societal cost, and their own efficacy; shutting down large congregations, closing bars, schools, or other venues, and aggressive social distancing. Choosing measures with the highest efficacy and lowest societal cost should have the highest priority, and there is a natural incentive to pick those measures.
Moreover, when there are only a few new cases per day, it is possible to prevent further spreading by TTI of individual cases~\cite{WHO2020}. This is a relatively cheap way to stop the virus from spreading. At higher levels of spread, one must rely on other, more expensive, region-wide measures, like closing down venues and stricter social distancing. Therefore, the cost per case is also lower when you can do TTI and contain the virus entirely. 

We stress the point where a transition is made from `firefighting' (TTI) to society wide restrictions, which happens at a few cases.  A few cases means all contacts can be traced, quarantined, and further spread can effectively be prevented with other people having to make only relatively small changes to their daily lives. After this point, more drastic society-wide measures will be required. However, a similar mechanism occurs at higher levels of circulation. If some number of measures together with TTI is sufficient to prevent further spreading, there is a similar transition point where TTI breaks down. Here again, when the limits of TTI are reached and the system loses some of its efficacy, harsher measures are needed to stop the virus from spreading further. This means that around points where TTI breaks down, the costs are locally convex, and it is important to protect the coverage of TTI at any level of circulation.
\bigskip

\begin{tcolorbox}
\textbf{A note on convexity of costs}
\bigskip

It immediately follows from a linear impact analysis that the larger the number of cases the larger the cost to society from an emergent disease. As TTI is a relatively cheap way to suppress the virus, the cost per case of suppressing the virus is lower when doing TTI than when not doing TTI. This means that around points where TTI breaks down and society must rely on additional measures, independent of what additional measures were already in place, the cost of suppressing the virus is locally convex. To further illustrate this point, consider two regions with small outbreaks that can both only just be effectively controlled through TTI. This requires a serious effort to prevent the outbreaks from creating community transmission and requireing more extensive measures. However, if the same number of cases would appear in a single outbreak the contact tracing system may pass the transition point and lose the completeness of its coverage. This is much more serious and costly for the affected region. 

Though the cost is increasing in the number of cases and has pockets of convexity around points where TTI breaks down, the cost as a function of the number of cases is not convex overall. The consequences for the healthcare system, society and the wider economy may still have  convex costs associated with them that are not considered here. Indeed, in general we see that as the shock to a complex system is larger, it is more costly or difficult to adapt to it. While a linear approximation may hold for a low level of outbreak, those costs may scale in a nonlinear way.

\end{tcolorbox}

\section{One Region, only domestic transmissions}

Here we abstract from considering inter-regional travel. Considering only the cost of halting local transmission, we find the optimal solution is to go to zero cases.

{\itshape  1. \ \ If a region is willing to implement measures to prevent spread at one level in number of cases, it must also be willing to prevent the spread at a lower level, since it will be cheaper to do so and has only positive other effects.}
\bigskip

Relaxing measures to save costs while increasing case numbers, only to implement more stringent measures later to reduce the virus presence, is more costly than keeping measures at such a level that case numbers always keep going down. This follows from the fact that costs are an increasing function of case numbers. As an example, recent increases in numbers of cases after relaxing containment measures and travel restrictions in Europe, followed by new travel restrictions and stronger containment measures within countries indicates policy choices are non-optimal. From the first proposition quickly follows a second one:
\bigskip

{\itshape 2. \ \ Without imported cases, the cheapest policy is to have zero cases.}
\bigskip

When there are no imported cases, the increasing nature of the cost function tell us that the total cost over time is lowest when there are no new cases. This implies that the best policy is at either 0 cases or by having no policies at all. For COVID, having no policies at all would mean that a majority would be infected, suffering damages and economic costs that should certainly be avoided. Some of the economic harm occurs directly because of the health harm, but also from risk reduction by individuals regardless of government policy. This spontaneous adoption of a social policy as the number of transmission events increases leads to high economic costs. 

We note that some argue that it would impossible to go to zero cases~\cite{Tegnell2020}, but despite our best efforts we found no reason for this to be the case. Instead, empirical evidence suggests not only that it is possible to go to zero cases, but also that it is less costly than controlling the spread at a higher level. The intuitive result that there is no trade off between the economy and fighting the virus also follows from empirical data, as is shown in~\cite{OWID2020c}. Comparing the COVID-19 death rate with the latest GDP data, we in fact see the opposite: countries that have managed to protect their population’s health in the pandemic have generally also protected their economy too.

\section{Policy for one region, including imported cases}

Here we consider the costs of suppressing local transmission and the cost of limiting imported cases from the perspective of an individual region without
policy cooperation with the rest of the world. We find that the optimal situation is with a very low number of cases. 
\bigskip

From the perspective of any individual region, the costs of border controls that are needed to limit imported cases are decreasing in the number of newly imported cases as long as the virus is present in the world. Costs include the loss of economic benefits of the travel in tourism and other business activity. These border control and other economic costs are offset by the cost of the outbreak response (TTI) or wider social restrictions that are needed to control transmission due to imported cases when travel is present. In this paper we deal only with the following costs; the cost of limiting local transmissions $c_T(x)$ to suppress the virus when there are $x$ new infections each day, the cost of border controls $c_b(I)$ with $I$ the average number of infected cases per day, and the cost that the outbreak brings to society $c_O(x)$:

$$cost = c_T + c_b - c_O.$$

The cost of the outbreak $c_O$ is written down with a minus as we consider the holistic cost of the policies to prevent spread. This means that while the cost of implementing the policy increases as we put in more effort to control this disease, the reduction in the effects of the outbreak and its associated costs will mean the total cost will be substantially reduced. $c_T(x)$ and $c_O(x)$ are both increasing in the number of cases as we shall see next, and $c_b(I)$ is decreasing in the average number of imported cases (thus increasing with stronger border control). The cost of the outbreak $c_O$ pertains to health effects, indirect economic effects, restricted liberties, and uncertainties to arise from a virus outbreak.

Suppose a traveler coming from any other region $R$  at a certain moment in time is infectious with probability $p_R$. The expected value of the number of imported cases is then  $\sum_R p_R k_R=I$, where $k_R$ is the number of travelers from region $R$. Since the probability $p_R$ is different for each region, it is more effective to prevent imported cases from regions with high probability of importing infectious travelers. The cost of accepting a low number of cases, for instance by allowing only travel from areas with very low infection rates is thus lower per avoided case by limiting travel from high-risk areas. Similarly, screening arriving passengers with temperature checks, testing and/or imposing quarantines can be relatively cheap ways to limit the amount of imported virus.
\bigskip

{\itshape  3. \ \ The cost of controlling the virus is minimal for some constant, possibly fractional, positive case number.}
\bigskip

In this case the cost function for limiting the virus spread changes from when there is only domestic transmission. The number of new domestic infections $x$ with associated policy cost $c_T(x)$ is insufficient to suppress the virus when there are also imported cases. Instead cost of policy to suppress the virus $c_T(x')$ must be made with $x'=x+\alpha I$, where $\alpha I$ is the number of imported cases plus additional cases of transmission resulting from those imported cases, which we define to be given by a constant multiple of $I$. Additionally, there are costs of border control $c_b(I)$. The policymaker may then solve the sequential minimization problem

$$\min_{I} \{\  \min_{x} [ c_T(x+\alpha I)+c_b(I) ] \ \} = \min_I [c_T(\alpha I)+c_b(I) ]$$
$$\equiv \min_I C(I),$$
where the cost aggregate $C(I)$ is the minimal cost of policy to suppress the virus given any level of  domestic transmissions $x$ each day including imported cases. The minimization is at $x=0$ as $c_T(x)$ is an increasing function of $x$. Therefore, assuming the cost of total and complete travel restriction is high enough, $C(I)$ is decreasing at $I=0$ when $c_b$ dominates, and increasing as $I$ is large enough (for sufficiently large $I$, $c_T$ dominates $c_b$).  $c_b(I)$ is not just decreasing in the number of imported cases, it is also true that the most expensive travel restrictions are for the most essential travelers. So, it is easier to prevent $\it{some}$ imported cases (compared to no border controls at all), than to have no imported cases at all (compared to allowing a low number of travelers with risk of importing $\it{some}$ cases).

The assumptions on costs associated with 
restrictions
limiting 
travel leads to a conclusion that there is an optimal number of imported cases. However, we note that the minimum may occur when the number of imported cases $I$ is so low that it is fractional, i.e. for a rate of importing cases that is less than one (so, in practice intermittently periods of nonzero cases). Importing new cases brings the risk of causing a significant initially undetected outbreak requiring both extensive TTI and social restrictions, as in recent experience in New Zealand~\cite{lewis2020}. Since these costs are large compared to conventional tourism and business opportunities, this is very well possible. Careful scrutiny of the associated costs and benefits of travelers in terms of the overall costs is well justified.

We note also that the cost of allowing imported cases varies between countries. For example, in island nations, controlling borders is easier. Conversely, for economies that heavily rely on tourism the benefit of tourism and thus the cost of closing borders is higher relative to the domestic economy~\cite{oecd2020}. Where cross-border travel plays a larger role in the economy, it becomes more expensive and difficult to implement border restrictions. This is also related to the size of a region. The effect is  reminiscent to how in physics large objects have a low surface-area to volume ratio. Similarly, large countries have a lower border cost to economy ratio for restricting cross-border travel compared to small countries.

Thus, for example, it would be expected that Luxembourg would be more dependent on allowing cross border travel than Germany. However, we note that Luxembourg imposed travel restrictions that allowed cross border workers while not allowing tourism. For this and other reasons found in for example~\cite{shen2020}, we can infer that restricting cross border workers would be a higher cost, and that workers would also represent a lower transmission risk either due to inherent social, psychological, and behavioral differences from tourists and/or due to enhanced precautions that they would take due to the conditions present. Moreover, it is well known that the smallest European countries, including Andorra, Luxembourg, Liechtenstein and Monaco, 
were initially
among the best performers in reducing COVID cases in Europe~\cite{OWID2020} 
(however, when European travel restrictions were collectively removed their situation worsened). 
The reason is not clear from this economic analysis but may be due to the possibility that smaller countries are more nimble, in particular having a greater ability to act collectively in response to the outbreak and optimize policy decisions. Some support for this observation arises from the recognition in larger countries such as Germany that differentiated policy decisions in localities allows better optimization of the response efforts~\cite{OWID2020b}.

It is important to recognize that using a reference for the benefits of travel that is in a COVID-free (pre-COVID) world is the wrong starting point for calculating the cost impacts of travel restrictions. Instead we need to consider the overall costs of travel limitations. Thus, for example, when a country calculates the benefit of allowing travel because of a per traveler tourism income in the pre-COVID world, this is not an estimate of the benefit in a world with COVID. Without travelers a country would have a different level of outbreak response (TTI) and, significantly, a different need for widespread social restrictions affecting economic activity~\cite{FT2020,McKinsey2020}. Due to the high costs of outbreak control measures, the number of travelers that provides a benefit is much lower than those in a world without COVID, and the desirable/optimal number decreases with increasing prevalence of COVID in other locations. This is in line with other analyses~\cite{Green2020,norman2020}.

\section{Policies for two regions}

Here we consider both local transmissions and imported cases, in a situation where two regions consider each other's policies and disease dynamics. We find that cooperation and coordination can play an important role in getting to low circulation.

Suppose there are only two regions, $i$ and $j$, both in which a percentage of  $L_\nu$ people are infected, $\nu \in\{i,j\}$. We assume that each day $k$ randomly selected\footnote{In reality, those who travel more often may well not be a random selection. We abstract from that for simplicity, and note this is an underestimation of the number of imported cases.} people travel from $i$ to $j$, and $N_\nu$ is the population of each region. The probability of having at least $n$ cases in those $k$ people is given by
$$P(\#imported\ cases \leq n) = \sum_{\nu=1}^n \frac{\binom{L_i N_i}{\nu}\binom{N_i -L_i N_i}{k-\nu} }{\binom{N}{k}}, $$
which approaches 
$$\sum_{\nu=1}^n \binom{k}{\nu} L_i^\nu$$
when $k\ll N_i L_i\ll N_i$.\footnote{\scriptsize{
Notice that
$$\frac{\binom{L_i N_i}{\nu}\binom{N_i -L_i N_i}{k-\nu} }{\binom{N}{k}}= \frac{\binom{k}{\nu} \binom{N_i-k}{L_i N_i -\nu}}{\binom{N_i}{N_i L_i}}=$$ 
$$= \binom{k}{\nu} \frac{(N_i-L_i N_i) ... (N_i-L_i N_i - (k-\nu)) \times (L_i N_i ... (L_i N_i - \nu))}{N_i ... (N_i - k)}\rightarrow$$
$$\binom{k}{\nu} L_i^\nu.$$
}}
The expected number of imported cases from region $i$ to region $j$ given $k$ travelers is given by 

$$\sum_{\nu=1}^k \binom{k}{\nu} L_i^\nu \nu\equiv I_{ik}$$
Region $i$ can take action and through screening or reducing travel get $I_{ik}$ down to $I_{ik}F$. When region $j$ assumes that there are cases in region $i$ and there will be cases indefinitely, it will weigh the cost of limiting travel from region $i$ against the cost of accepting a number of imported cases from region $i$. However, when the two regions consider their strategy holistically, they would consider themselves one region and would go for zero cases, as discussed above. This would require stronger measures to fight the epidemic in the region with more active transmission as well as the region with less transmission, and/or stricter limitations on the number of people traveling from the region with more active transmission. The subsequent result would be lower costs for both.

We can formalize the cost for region $j$ similar to the one-region case. The cost of limiting domestic cases $c_T(x)$ to a level of $x$ cases per day is represented by $c_T(0)=c_0>0, c_T'(x)>0$ (increasing). The cost of limiting imported cases to a level of $y$ cases, represented by $c_b(x)$, per day is characterized by $c_b(0)>0, c_b'(y)<0$, (decreasing), $c_b(I_{ik})=0$ (no border costs when fully open). Region $j$ does not consider the impact of its own number of cases on the strategy in region $i$. In this scenario we identify the minimization problem for region $j$ as
\begin{equation}
\label{eqn:minimization}
\min_{F} c_b + c_T
\end{equation}with interior solution 
$$\frac{d c_T(x+\alpha I_{ik} F)}{dF}= - \frac{d c_b(I_{ik} F)}{dF}.$$
The minimization problem would have boundary solution $F=1$ (no border control at all) when $\frac{d c_T(x+\alpha I_{ik}F)}{dF}|_{F=1}<-\frac{d c_b(I_{ik}F)}{dF}|_{F=1}$. This is only a hypothetical scenario, as it occurs when the cost of limiting imported cases is so expensive that it is always cheaper to just accept the cases and limit the spread locally. Additionally, the boundary solution for $F=0$ is found when the condition holds that the cost of holding cases at zero, plus the cost of holding cases at near-zero is when a new case is imported, is larger than the marginal cost savings of allowing even a little bit of travel between regions $i$ and $j$: $\frac{d c_T(\alpha I_{ik}F)}{dF}|_{F=0}>-\frac{d c_b(I_{ik}F)}{dF}|_{F=0}$.\footnote{When all borders are completely closed, the pandemic preparedness at 0 zero cases can also be reduced significantly, giving rise to the new assumptions $\lim_{x\downarrow 0} c_T(x) = c_0$ and $c_T(0)\approx 0$, so that we alternatively find $F=0$ when $c_0+\frac{d c_T(\alpha I_{ik}F)}{dF}|_{F=0}>-\frac{d c_b(I_{ik}F)}{dF}|_{F=0}.$}

The above analysis gives rise to somewhat of a paradox: If region $i$ acts assuming region $j$ does not eradicate the virus, and region $j$ assumes $i$ will not eradicate the virus, then for both regions the lowest-cost approach is to accept some infected individuals (perhaps fractional per day, i.e. some risk of an infected individual arriving). When both region $i$ and $j$ determine their strategy together, they are back in the situation of only domestic transmission described above, and they should eradicate the virus. The maximization problem for a single region is the same as in section B above with one region without imported cases. The cost here reduces back to
$$cost = c_T (x) - c_O(x)$$
and since $c_T$ is increasing in $x$ and $-c_O$ is decreasing in $x$, the cost is minimal when the number of cases is zero. This occurs when both regions act under one government. This is naturally also the case when both regions coordinate. When weaker assumptions are made, zero can still be the optimal strategy. If one region assumes rationality on the part of the other region, then its own rational solution is to aim for zero cases. 
In the case where one or both regions assume, wrongly or not, that the other region will not contain its virus presence, it faces the choice of limiting transport between regions indefinitely and at high cost, and may as a consequence accept some transmission.

So, while even without coordination each region should aim for zero cases when assuming rationality on the part of the other region, this still highlights the utility of inter-regional cooperation. A practical example of where such a coordinated effort is possible, is found in the European Union, where countries have recently started a shared response~\cite{EU2020}. As an extension of this analysis, it has been shown that due to the fact that Europe did its lockdowns in its first wave at the same time, they were much more effective \cite{ruktanonchai2020}.

One reason a region may not go to zero is because it makes the assumption that it's impossible to do so. Although some scientists~\cite{Tegnell2020} suggest that it is impossible to go to zero cases, there is no reason for that assumption. In fact, several countries have already shown it can be done. Finally, we note that at the heart of this there is a collective action problem regarding the cost of limiting transportation; if both regions go to zero cases, travel between the regions could be freed up and corona-free without any cost, and that would be the optimal (lowest-cost) strategy for each region. If either region decides not to, it is harder for the other region to likewise go to zero because it has to maintain travel restrictions indefinitely. The solution to the collective action problem in this case is simple, as it is in each region's individual interest to have a low number of cases.  There is a tragedy of the commons dynamics due to the fact that a region maintaining a large number of cases has the externality of infecting other regions. What is unusual in comparison to typical analyses, is that what seems to be an exploitative action by one of the players is actually a cost increasing (irrational) action.

Note that we have not used $c_O$ much in this analysis. In addition to the direct cost of policy discussed here, there are many more arguments to contain outbreaks. These pertain to the costs associated with $c_O$; health effects (many people get sick, and even die), indirect economic effects (even without strong anti-transmission measures, the economy would respond strongly to the virus presence \cite{koolman2020}), ongoing restricted liberties, ethics, possible mutations and uncertainties \cite{norman2020} that are exacerbated by not going to zero cases. The costs included in $c_O$ could be absorbed into the economic costs of suppressing the virus $c_T$, and the analysis would not change qualitatively.

In conclusion, we identified a reason for a nonzero COVID outcome as the individual country solution to a cost-of-policy optimization where it is assumed that other countries themselves irrationally adopt a costly nonzero COVID policy. This response is one that some European countries seem to be aiming for given their policies are sufficient to keep outbreaks in check, but insufficient to eliminate the virus from their countries entirely. We note, however, that optimization indicates that if a region is willing to implement measures to prevent spread at one level in number of cases, it must also be willing to prevent the spread at a lower level, since it will be cheaper to do so and has only positive other effects. Therefore the strategy of keeping the virus in check is at higher levels of circulation is not only impractical due to the many uncertainties surrounding the further development of the virus and its spread, but also not optimal from a dynamic optimization perspective. Moreover, we note that travel induced cases should be limited to very low levels due to the very high costs of country wide actions to contain outbreaks in the presence of COVID. Lastly, we note that cooperation (and/or rationality assumptions) should be promoted to give rise to joint zero COVID strategies.


    

\bibliographystyle{plainnat}
\bibliography{bibliography}

\begin{thebibliography}{24}
\providecommand{\natexlab}[1]{#1}
\providecommand{\url}[1]{\texttt{#1}}
\expandafter\ifx\csname urlstyle\endcsname\relax
  \providecommand{\doi}[1]{doi: #1}\else
  \providecommand{\doi}{doi: \begingroup \urlstyle{rm}\Url}\fi

\bibitem[Acemoglu et~al.(2020)Acemoglu, Chernozhukov, Werning, and
  Whinston]{acemoglu2020}
Daron Acemoglu, Victor Chernozhukov, Iv{\'a}n Werning, and Michael~D Whinston.
\newblock A multi-risk sir model with optimally targeted lockdown.
\newblock Technical report, National Bureau of Economic Research, 2020.

\bibitem[Antr{\`a}s et~al.(2020)Antr{\`a}s, Redding, and
  Rossi-Hansberg]{antras2020}
Pol Antr{\`a}s, Stephen~J Redding, and Esteban Rossi-Hansberg.
\newblock Globalization and pandemics.
\newblock Technical report, Harvard University Working Paper, 2020.

\bibitem[Bar-Yam(2016)]{baryam2016}
Yaneer Bar-Yam.
\newblock Transition to extinction: Pandemics in a connected world, 2016.

\bibitem[Bognanni~M and K(2020)]{bognana2020}
Kolliner~D Bognanni~M, Hanley~D and Mitman K.
\newblock Economics and epidemics: Evidence from an estimated spatial econ-sir
  model, 2020.

\bibitem[Green et~al.(2020)Green, Shen, and Bar-Yam]{Green2020}
Aaron Green, Chen Shen, and Yaneer Bar-Yam.
\newblock Case studies of covid-19 travel restrictions.
\newblock \emph{New England Complex Systems Institute}, 2020.

\bibitem[in~Data(2020{\natexlab{a}})]{OWID2020}
Our~World in~Data.
\newblock Luxembourg: Coronavirus pandemic country profile.
\newblock \url{https://ourworldindata.org/coronavirus/country/luxembourg},
  2020{\natexlab{a}}.
\newblock Accessed: 2020-09-11.

\bibitem[in~Data(2020{\natexlab{b}})]{OWID2020b}
Our~World in~Data.
\newblock Emerging covid-19 success story: Germany’s strong enabling
  environment.
\newblock \url{https://ourworldindata.org/covid-exemplar-germany},
  2020{\natexlab{b}}.
\newblock Accessed:2020-09-11.

\bibitem[in~Data(2020{\natexlab{c}})]{OWID2020c}
Our~World in~Data.
\newblock Which countries have protected both health and the economy in the
  pandemic?
\newblock \url{https://ourworldindata.org/covid-health-economy},
  2020{\natexlab{c}}.
\newblock Accessed:2020-09-11.

\bibitem[Kissler et~al.(2020)Kissler, Tedijanto, Goldstein, Grad, and
  Lipsitch]{kissler2020}
Stephen~M Kissler, Christine Tedijanto, Edward Goldstein, Yonatan~H Grad, and
  Marc Lipsitch.
\newblock Projecting the transmission dynamics of sars-cov-2 through the
  postpandemic period.
\newblock \emph{Science}, 368\penalty0 (6493):\penalty0 860--868, 2020.

\bibitem[Koolman(2020)]{koolman2020}
Xander Koolman.
\newblock It's the virus stupid!
\newblock \emph{Academic Health Economists' Blog}, 2020.

\bibitem[Kulldorf~M(2020)]{GBD2020}
et~al. Kulldorf~M.
\newblock The great barrington declaration.
\newblock \url{https://gbdeclaration.org/}, 2020.

\bibitem[Lewis(2020)]{lewis2020}
Dyani Lewis.
\newblock "we felt we had beaten it": New zealand's race to eliminate the
  coronavirus again.
\newblock \emph{Nature}, 584\penalty0 (7821):\penalty0 336, 2020.

\bibitem[McKinsey(2020)]{McKinsey2020}
McKinsey.
\newblock Covid-19: Implications for business.
\newblock
  \url{https://www.mckinsey.com/business-functions/risk/our-insights/covid-19-implications-for-business},
  2020.

\bibitem[Norman et~al.(2020)Norman, Bar-Yam, and Taleb]{norman2020}
Joseph Norman, Yaneer Bar-Yam, and Nassim~Nicholas Taleb.
\newblock Systemic risk of pandemic via novel pathogens—coronavirus: A note.
\newblock \emph{New England Complex Systems Institute (January 26)}, 2020.

\bibitem[OECD(2020)]{oecd2020}
OECD.
\newblock Tourism policy responses to the coronavirus (covid-19).
\newblock \url{https://www.oecd.org/coronavirus/}, 2020.
\newblock Accessed:2020-09-11.

\bibitem[Paterlini(2020)]{Tegnell2020}
Marta Paterlini.
\newblock "closing borders is ridiculous”: the epidemiologist behind
  sweden’s controversial coronavirus strategy.
\newblock \url{https://www.nature.com/articles/d41586-020-01098-x}, 2020.

\bibitem[Petersen et~al.(2018)Petersen, Petrosillo, Koopmans, Beeching,
  Di~Caro, Gkrania-Klotsas, Kantele, Kohlmann, Lim, Markotic,
  et~al.]{petersen2018}
Eskild Petersen, Nicola Petrosillo, Marion Koopmans, N~Beeching, A~Di~Caro,
  E~Gkrania-Klotsas, A~Kantele, R~Kohlmann, P-L Lim, A~Markotic, et~al.
\newblock Emerging infections—an increasingly important topic: review by the
  emerging infections task force.
\newblock \emph{Clinical Microbiology and Infection}, 24\penalty0 (4):\penalty0
  369--375, 2018.

\bibitem[Rijksoverheid(2020)]{rijksoverheid2020}
Rijksoverheid.
\newblock Tv-toespraak van minister-president mark rutte.
\newblock
  \url{www.rijksoverheid.nl/documenten/toespraken/2020/03/16/tv-toespraak-van-minister-president-mark-rutte},
  2020.

\bibitem[Ruktanonchai et~al.(2020)Ruktanonchai, Floyd, Lai, Ruktanonchai,
  Sadilek, Rente-Lourenco, Ben, Carioli, Gwinn, Steele,
  et~al.]{ruktanonchai2020}
Nick~Warren Ruktanonchai, JR~Floyd, Shengjie Lai, Corrine~Warren Ruktanonchai,
  Adam Sadilek, Pedro Rente-Lourenco, Xue Ben, Alessandra Carioli, Joshua
  Gwinn, JE~Steele, et~al.
\newblock Assessing the impact of coordinated covid-19 exit strategies across
  europe.
\newblock \emph{Science}, 369\penalty0 (6510):\penalty0 1465--1470, 2020.

\bibitem[Shen et~al.(2020)Shen, Killeen, Staines, and Bar-Yam]{shen2020}
Chen Shen, Gerry~F Killeen, Anthony Staines, and Yaneer Bar-Yam.
\newblock A green zone strategy for ireland, 2020.

\bibitem[Siegenfeld and Bar-Yam(2020)]{siegenfeld2020}
Alexander~F Siegenfeld and Yaneer Bar-Yam.
\newblock Eliminating covid-19: The impact of travel and timing.
\newblock \emph{arXiv preprint arXiv:2003.10086}, 2020.

\bibitem[Times(2020)]{FT2020}
Financial Times.
\newblock Real-time data show virus hit to global economic activity.
\newblock
  \url{https://www.ft.com/content/d184fa0a-6904-11ea-800d-da70cff6e4d3}, 2020.

\bibitem[Union(2020)]{EU2020}
European Union.
\newblock The common eu response to covid-19.
\newblock \url{https://europa.eu/european-union/coronavirus-response_en}, 2020.

\bibitem[WHO(2020)]{WHO2020}
WHO.
\newblock Director-general's opening remarks at the media briefing on covid-19
  - 16 march 2020.
\newblock \url{https://www.who.int/}, 2020.
\newblock Accessed: 2020-09-11.

\end{thebibliography}

\end{document}